\documentclass[aps, prl,twocolumn,reprint,superscriptaddress]{revtex4-2}
\usepackage{comment}
\usepackage{graphicx}
\usepackage{amsmath}
\usepackage{setspace}
\usepackage{braket}
\usepackage{mathtools}
\usepackage{siunitx}

\usepackage{threeparttable}
\usepackage{booktabs}
\usepackage{transparent}
\usepackage{orcidlink}
\usepackage{cleveref}
\usepackage{natbib,mathrsfs,accents}
\usepackage{siunitx}

\usepackage[caption=false]{subfig}
\usepackage{float}

\definecolor{seaborn1}{rgb}{0.00392156862745098, 0.45098039215686275, 0.6980392156862745}
\definecolor{seaborn2}{rgb}{0.8705882352941177, 0.5607843137254902, 0.0196078431372549}
\definecolor{seaborn3}{rgb}{0.00784313725490196, 0.6196078431372549, 0.45098039215686275}
\definecolor{seaborn4}{rgb}{0.8352941176470589, 0.3686274509803922, 0.0}
\definecolor{seaborn5}{rgb}{0.8, 0.47058823529411764, 0.7372549019607844}
\definecolor{seaborn6}{rgb}{0.5803921568627451, 0.5803921568627451, 0.5803921568627451}
\definecolor{seaborn7}{rgb}{0.9254901960784314, 0.8823529411764706, 0.2}
\definecolor{seaborn8}{rgb}{0.33725490196078434, 0.7058823529411765, 0.9137254901960784}

\newcommand\scalemath[2]{\scalebox{#1}{\mbox{\ensuremath{\displaystyle #2}}}}
\usepackage{url}

\usepackage[nohyperlinks, nolist]{acronym}
\newacro{BEC}[BEC]{Bose-Einstein condensate}
\newacro{DKC}[DKC]{delta-kick collimation}

\begin{document}

\title{Proposal for a Bose-Einstein condensate based test of Born’s rule using light-pulse atom interferometry}

\author{Simon Kanthak\,\orcidlink{0000-0002-6576-8753}}
\email[]{simon.kanthak@physik.hu-berlin.de}
\affiliation{Institut f\"{u}r Physik and IRIS Adlershof, Humboldt-Universit\"{a}t zu Berlin, Newtonstr. 15, 12489 Berlin, Germany}

\author{Julia Pahl\,\orcidlink{0009-0004-6379-0967}}
\affiliation{Institut f\"{u}r Physik and IRIS Adlershof, Humboldt-Universit\"{a}t zu Berlin, Newtonstr. 15, 12489 Berlin, Germany}

\author{Daniel Reiche\,\orcidlink{0000-0002-6788-9794}}
\affiliation{Institut f\"{u}r Physik and IRIS Adlershof, Humboldt-Universit\"{a}t zu Berlin, Newtonstr. 15, 12489 Berlin, Germany}

\author{Markus Krutzik} 
\affiliation{Institut f\"{u}r Physik and IRIS Adlershof, Humboldt-Universit\"{a}t zu Berlin, Newtonstr. 15, 12489 Berlin, Germany}
\affiliation{Ferdinand-Braun-Institut (FBH), Gustav-Kirchoff-Str.4, 12489 Berlin, Germany}

\begin{abstract}
We propose and numerically benchmark light-pulse atom interferometry with ultra-cold quantum gases as a platform to test the modulo-square hypothesis of Born's rule. Our interferometric protocol is based on a combination of double Bragg and single Raman diffraction to induce multipath interference in \acp{BEC} and block selected interferometer paths, respectively. In contrast to previous tests employing macroscopic material slits and blocking masks, optical diffraction lattices provide a high degree of control and avoid possible systematic errors like geometrical inaccuracies from manufacturing processes. In addition, sub-recoil expansion rates of delta-kick collimated \acp{BEC} allow to prepare, distinguish and selectively address the external momentum states of the atoms. This further displays in close-to-unity diffraction fidelities favorable for both high-contrast interferometry and high extinction of the blocking masks. In return, non-linear phase shifts caused by repulsive atom-atom interactions need to be taken into account, which we fully reflect in our numerical simulations of the multipath interferometer.  Assuming that the modulo-square rule holds, we examine the impact of experimental uncertainties in accordance with conventional \ac{BEC} interferometer to provide an upper bound of $5.7\times10^{-3}$ $\left(1.8\times10^{-3}\right)$ on the statistical deviation of $100$ $\left(1000\right)$ iterations for a hypothetical third-order interference term.
\end{abstract}


\maketitle

\section{Introduction}

It is the abstract beauty of modern theories that the deterministic dynamics of fields is computed on a hidden, often linear layer, while the actual observables can only be derived using squares or products of such fields. 
The translation between the ``abstract mathematical world of the first
layer [and] the concrete mechanical world of the second layer'' \cite{dyson07} is made by predefined rules.
Such rules can usually not be derived from first principles and are hence deeply connected to interpretations of the underlying theory.
This is perhaps most famously represented by the Born rule in quantum mechanics: It relates the measurement probabilities to the square moduli of the wave function's scalar product \cite{born26a}. 
On the one hand, testing the Born rule touches the fundamental core of quantum theory (see, e.g., Refs.~\cite{zurek03a,schlosshauer05,wang20,vaidman20a} for a recent discussion) and, in particular, the collapse of the wave function during the measurement process \cite{bassi13,mertens21}.  
On the other hand, Born's rule is the microscopic explanation for (quantum) superposition and interference phenomena and therefore plays a crucial role in the development of modern quantum technologies \cite{dowling03,wang19,amico22,hochrainer22}.

It may hence be no surprise that addressing the Born rule experimentally on a multitude of platforms has attracted substantial attention in the past decades \cite{arndt14,hochrainer22}.
To name only a few examples, single-particle interference was tested in a three-slit setup using photons \cite{sinha10}, large molecules \cite{cotter17}, or metastable helium atoms diffracted by an array of slits \cite{barnea18}. 
Alternatively, there are interference experiments with many particles and multiple slits \cite{pleinert20,pleinert21}, setups using half-spin registers in liquid state nuclear magnetic resonance \cite{park12}, and proposals to test Born's rule by means of a quantum computer with superconducting qubits \cite{sadana22}. 
So far, without finding any significant deviation from the theoretical expectation. 

Yet another, more practical, translation issue occurs when the concrete experimental reality is mapped to the theoretical model. 
Deviations from Born's rule can be rigorously defined using a so-called Sorkin-like parameter $\epsilon$ \cite{sorkin94} on the level of quantum states which should vanish, if Born's rule holds.
However, a one-to-one correspondence of such states with experimental setups can be misleading. 
For instance, in the conventional three-slit experiment (see, e.g., \cite{sinha10}), it has been shown that diffraction and light-matter interactions \cite{raedt12,sawant14,sinha15}, the inclusion of non-classical paths \cite{quach17} or an incorrect adjusting of photon numbers in the comparison of different setups \cite{skagerstam18} can lead to a non-zero Sorkin parameter. 
This does not indicate a violation of Born's rule, but rather an incomplete theoretical description.

A not yet considered platform suited for interferometric measurements is cold atoms \cite{chu85,kasevich92}, which constitute a high-sensitivity matter wave source for atom interferometry. While pioneering works employed material slits \cite{mlynek91} and gratings \cite{keith91}, optical lattices have proven to be the most promising technique due to the high degree of control and coherence of the diffraction process \cite{riehle91,kasevich91}. Three-path atom interferometers have already been employed, i.e. as a magnetometer \cite{hinderthur97} or to measure the fine-structure constant \cite{plotkin18}. In this proposal, such a three-path interferometer is studied for a new type of Born rule test. For this, quantum degenerate gases are of particular interest due to their global phase coherence, however suffer from repulsive interaction which cause the phase distribution to evolve non-linear \cite{simsarin00}. In fact, since the Sorkin parameter is traditionally formulated for linear dynamics, such as the one-particle Schr\"odinger equation, deviations naturally occur when multi-particle and non-linear interactions are considered \cite{rozema21,namdar23}.
A non-zero Sorkin parameter is not sufficient to disprove Born's rule. It is instead the parameter that needs to be adjusted to account for nonlinearities of the system.\\

\begin{figure}[t!]
\begin{center}
\includegraphics[width=1\columnwidth]{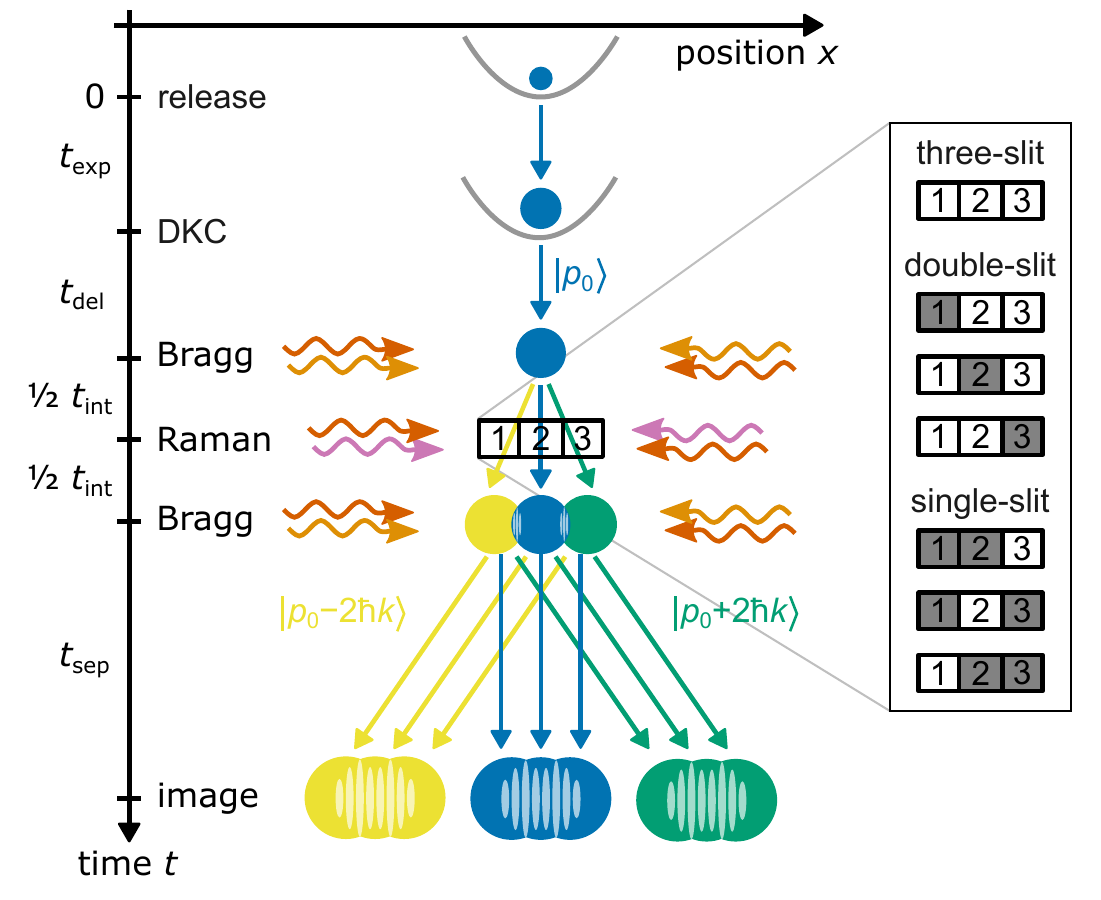}
\caption{Experimental protocol for a \ac{BEC} based test of Born's rule using light-pulse atom interferometry. The interferometric geometry consists of an Open-Ramsey like interferometer employing double Bragg diffraction to symmetrically split and superimpose matter waves. Single Raman pulses are interposed between the two Bragg pulses and serve as momentum selective blocking masks to form various slit configurations $\{\scalemath{0.95}{\fcolorbox{black}{white}{$1\vphantom{3}$}}, \scalemath{0.95}{\fcolorbox{black}{white}{$2$}}, \scalemath{0.95}{\fcolorbox{black}{white}{$3$}}\}$. Interference patterns form in three interferometer outputs, which are recorded for each of these configurations.}
\label{fig:scheme}
\end{center}
\end{figure}

In the present manuscript, we propose an experimental protocol for testing the modulo-square-rule via light-pulse atom interferometry using \acfp{BEC}. At sufficiently low temperatures, the atomic ensembles can effectively be described by a single wave function under the influence of non-linear self-interaction, i.e. the Gross-Pitaevskii equation \cite{fitzek20,kanthak21, burchianti20,yao22}. 
The square modulus of the wave function then corresponds to the spatial density distribution of the condensate which is directly experimentally accessible, e.g. by absorption or fluorescence imaging techniques. 
In this way, our proposal avoids common pitfalls of photon-based measurements, i.e. diffraction of light at mask interfaces, non-classical paths as well as cross-talks between interferometer paths or detector dark counts~\cite{soellner12}.
In contrast, we fully appreciate the nonlinearities in the evolution of the \ac{BEC} and compute a general (non-zero) Sorkin parameter for a specific experimental implementation under the assumption that the modulo-square rule holds. 
This \emph{implicitly and indirectly} assumes also the Born rule to be applicable, such that any experimental deviation from our computed Sorkin parameter could hint towards a dynamics that goes beyond standard quantum mechanics.

\section{Experimental protocol}

\begin{figure}[t!]
\begin{center}
\includegraphics[width=1\columnwidth]{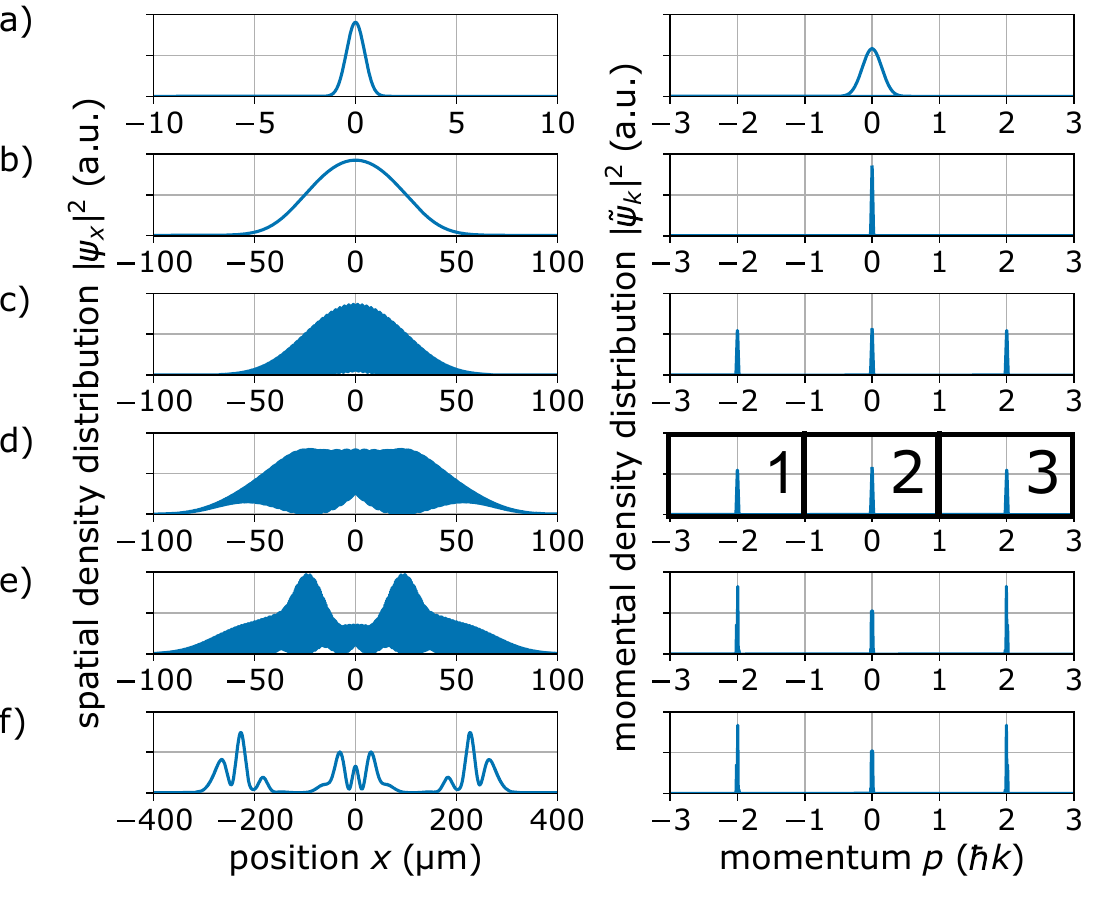}
\caption{Numerical simulation of the Open-Ramsey interferometer with blocking mask. Temporal evolution of the condensate's density distribution in position (left) and momentum space (right). A \ac{BEC} expands freely after a) release from the intial trap before being b) delta-kick collimated. Two subsequent double Bragg pulses c) split and e) redistribute the wave packets among three momentum states, which then form an interference pattern in f) the interferometer output ports. d) The three-slit mask \scalemath{0.95}{\fcolorbox{black}{white}{$1$}\fcolorbox{black}{white}{$2$}\fcolorbox{black}{white}{$3$}} is indicated in momentum space.}
\label{fig:simulation}
\end{center}
\end{figure}

The proposed experimental protocol is sketched in figure~\ref{fig:scheme} with the corresponding numerical evolution of the atomic density distribution shown in figure~\ref{fig:simulation}. We mimic the three-slit interferometric setup from the optical experiment \cite{sinha10} for matter waves, where certain paths can be blocked. The interferometer geometry consists of an Open-Ramsey like interferometer employing double Bragg diffraction for symmetric beam splitting and recombination of the matter-wave packets. Interposed, single Raman pulses serve as momentum selective blocking masks to form various slit configurations. 

Initially, a ground-state \ac{BEC} is generated inside and subsequently released from a harmonic potential. In practice, this can be realized by magnetic or optical fields. After freely expanding for $t=t_\text{exp}$, the \ac{BEC} is re-exposed to the same potential for a short duration of $\tau_\text{dkc}$ to narrow its momentum distribution via \ac{DKC} \cite{ammann97}. The \ac{BEC} then continues to expand freely  for $t=t_{\text{del}}$ in its zero-momentum state ($|p\rangle=|p_0\rangle$), which serves as the \emph{interferometer input port}.

We apply two consecutive light pulses for double Bragg diffraction at $t=t_{\text{exp}}+t_{\text{del}}$ and $t=t_{\text{exp}}+t_{\text{del}}+t_{\text{int}}$ \cite{giese13,giese15,ahlers16} to construct an Open-Ramsey interferometer. The duration and intensity of the pulses are chosen such that the wave packet at zero momentum is first split into three distinct momentum states $|p\rangle=\{|p_0 - 2\hbar k\rangle, |p_0\rangle, |p_0 + 2\hbar k\rangle\}$ of equal population, where $k$ is the wave vector of the light fields. Those are the \emph{interferometer paths} $\{\scalemath{0.9}{\fcolorbox{black}{white}{$1\vphantom{3}$}}, \scalemath{0.9}{\fcolorbox{black}{white}{$2$}}, \scalemath{0.9}{\fcolorbox{black}{white}{$3$}}\}$ which will be either open (\scalemath{0.9}{\fcolorbox{black}{white}{$\phantom{3}$}}) or blocked (\scalemath{0.9}{\fcolorbox{black}{gray}{$\phantom{3}$}}).  

The second pulse redistributes each of the wave packets among the three momentum states. These are the  \emph{interferometer output ports}, which become separable at $t=t_{\text{exp}}+t_{\text{del}}+t_{\text{int}}+t_{\text{sep}}$. The pulse separation of the interferometer $t_\text{int}$ is chosen such that the three wave packets in each of the ports spatially overlap and, due to their displacement of $t_\text{int}\cdot 2\hbar k/m$ and non-vanishing momentum width $\sigma_p$, form an interference pattern. Here, $m$ denotes the atomic mass. We remark that the number of visible interference fringes and their spacing depends on the momentum distribution and proper timing of the sequence \cite{miller05}.

Additionally, a combination of single Raman pulses at $t=t_{\text{exp}}+t_{\text{del}}+t_{\text{int}}/2$ is applied between the double Bragg pulses. In addition to an external momentum state change, the Raman pulses may address the internal state of the wave packets \cite{kasevich91,mueller08}. The duration and intensity are adjusted for a $\pi$-pulse such that the population is completely transferred between two long-lived states called $\ket{\text{g}_1}$ and $\ket{\text{g}_2}$, i.e. a ground-state manifold. The individual Doppler-shifts of the three momentum states allow to selectively address the wave packets in the \emph{interferometer paths}. Since the detection of the \emph{interferometer output} will only be sensitive to one internal state, the Raman pulses thus serve as momentum selective blocking masks. We utilize these to block certain paths in $3!+1$ experimental iterations - similarly to the optical counterpart of the experiment \cite{sinha10}.
 
At $t=t_{\text{exp}}+t_{\text{del}}+t_{\text{int}}+t_{\text{sep}}$, we determine the spatial distribution of the atomic density $\rho=|\psi_{x}|^2$ in the lower-energetic state of the ensemble for each configuration and define a Sorkin parameter \cite{sorkin94}
\begin{multline}
    \epsilon=\overbrace{\rho_{_{\,\mathrm{\scalemath{0.9}{  \fcolorbox{black}{white}{$_1$}\fcolorbox{black}{white}{$_2$}\fcolorbox{black}{white}{$_3$} }}}}}^{\text{three-slit}}-\overbrace{\rho_{_{\,\mathrm{\scalemath{0.9}{  \fcolorbox{black}{white}{$_1$}\fcolorbox{black}{white}{$_2$}\fcolorbox{black}{gray}{$_3$} }}}}-\rho_{_{\,\mathrm{\scalemath{0.9}{  \fcolorbox{black}{gray}{$_1$}\fcolorbox{black}{white}{$_2$}\fcolorbox{black}{white}{$_3$} }}}}-\rho_{_{\,\mathrm{\scalemath{0.9}{  \fcolorbox{black}{white}{$_1$}\fcolorbox{black}{gray}{$_2$}\fcolorbox{black}{white}{$_3$} }}}}}^{\text{two-slit}}\\+\underbrace{\rho_{_{\,\mathrm{\scalemath{0.9}{  \fcolorbox{black}{white}{$_1$}\fcolorbox{black}{gray}{$_2$}\fcolorbox{black}{gray}{$_3$} }}}}+\rho_{_{\,\mathrm{\scalemath{0.9}{  \fcolorbox{black}{gray}{$_1$}\fcolorbox{black}{white}{$_2$}\fcolorbox{black}{gray}{$_3$} }}}}+\rho_{_{\,\mathrm{\scalemath{0.9}{  \fcolorbox{black}{gray}{$_1$}\fcolorbox{black}{gray}{$_2$}\fcolorbox{black}{white}{$_3$} }}}}}_{\text{single-slit}} -\underbrace{\rho_{_{\,\mathrm{\scalemath{0.9}{  \fcolorbox{black}{gray}{$_1$}\fcolorbox{black}{gray}{$_2$}\fcolorbox{black}{gray}{$_3$} }}}}}_{\text{offset}}\text{.}
    \label{eq:sorkin_epsilon}
\end{multline}
The Sorkin parameter $\epsilon$ measures the validity of the modulo-square rule \cite{sorkin94}.
If the atomic gas can be assumed to be sufficiently dilute, self-interaction of the \ac{BEC} is negligible and $\epsilon=0$ at all times, provided the modulo-square rule holds.   
However, a parameter $\epsilon\neq0$ does not necessarily denote a violation of the rule (in contrast to pure quantum states \cite{sorkin94,pleinert20,pleinert22}), since already emerging self-interaction will naturally lead to a non-vanishing $\epsilon$. Further, experimental uncertainties (i.e. imperfect interferometer pulses, limited detection sensitivity) may increase the value of $\epsilon$. 

For scale invariance, we use a normalized variant of the Sorkin parameter as defined in Ref.~\cite{sinha10}
\begin{equation}
    \kappa = \frac{\epsilon}{\delta} \text{,}
    \label{eq:sorkin_kappa}
\end{equation}
where
\begin{multline}
    \delta=\left| \rho_{_{\,\mathrm{\scalemath{0.9}{  \fcolorbox{black}{white}{$_1$}\fcolorbox{black}{white}{$_2$}\fcolorbox{black}{gray}{$_3$} }}}}-\rho_{_{\,\mathrm{\scalemath{0.9}{  \fcolorbox{black}{white}{$_1$}\fcolorbox{black}{gray}{$_2$}\fcolorbox{black}{gray}{$_3$} }}}} -\rho_{_{\,\mathrm{\scalemath{0.9}{  \fcolorbox{black}{gray}{$_1$}\fcolorbox{black}{white}{$_2$}\fcolorbox{black}{gray}{$_3$} }}}}+\rho_{_{\,\mathrm{\scalemath{0.9}{  \fcolorbox{black}{gray}{$_1$}\fcolorbox{black}{gray}{$_2$}\fcolorbox{black}{gray}{$_3$} }}}} \right| +\\
    \left| \rho_{_{\,\mathrm{\scalemath{0.9}{  \fcolorbox{black}{gray}{$_1$}\fcolorbox{black}{white}{$_2$}\fcolorbox{black}{white}{$_3$} }}}}-\rho_{_{\,\mathrm{\scalemath{0.9}{  \fcolorbox{black}{gray}{$_1$}\fcolorbox{black}{white}{$_2$}\fcolorbox{black}{gray}{$_3$} }}}} -\rho_{_{\,\mathrm{\scalemath{0.9}{  \fcolorbox{black}{gray}{$_1$}\fcolorbox{black}{gray}{$_2$}\fcolorbox{black}{white}{$_3$} }}}}+\rho_{_{\,\mathrm{\scalemath{0.9}{  \fcolorbox{black}{gray}{$_1$}\fcolorbox{black}{gray}{$_2$}\fcolorbox{black}{gray}{$_3$} }}}} \right| +\\
    \left| \rho_{_{\,\mathrm{\scalemath{0.9}{  \fcolorbox{black}{white}{$_1$}\fcolorbox{black}{gray}{$_2$}\fcolorbox{black}{white}{$_3$} }}}}-\rho_{_{ \mathrm{\scalemath{0.9}{\,\fcolorbox{black}{white}{$_1$}\fcolorbox{black}{gray}{$_2$}\fcolorbox{black}{gray}{$_3$} }}}} -\rho_{_{\,\mathrm{\scalemath{0.9}{  \fcolorbox{black}{gray}{$_1$}\fcolorbox{black}{gray}{$_2$}\fcolorbox{black}{white}{$_3$} }}}}+\rho_{_{\,\mathrm{\scalemath{0.9}{  \fcolorbox{black}{gray}{$_1$}\fcolorbox{black}{gray}{$_2$}\fcolorbox{black}{gray}{$_3$} }}}} \right|
    \label{eq:sorkin_delta}
\end{multline}
denotes the expected two-path interference. In order to benchmark our protocol, we want to put the systematics of an experimental realization on a scale for an unexpected deviation from Born's rule. The unitless quantity $\kappa$ is independent of the specific experimental implementation, and thus allows for comparison with previously published works \cite{sinha10,cotter17,barnea18,soellner12}. 

\begin{figure*}[t]
\includegraphics[width=2.05\columnwidth]{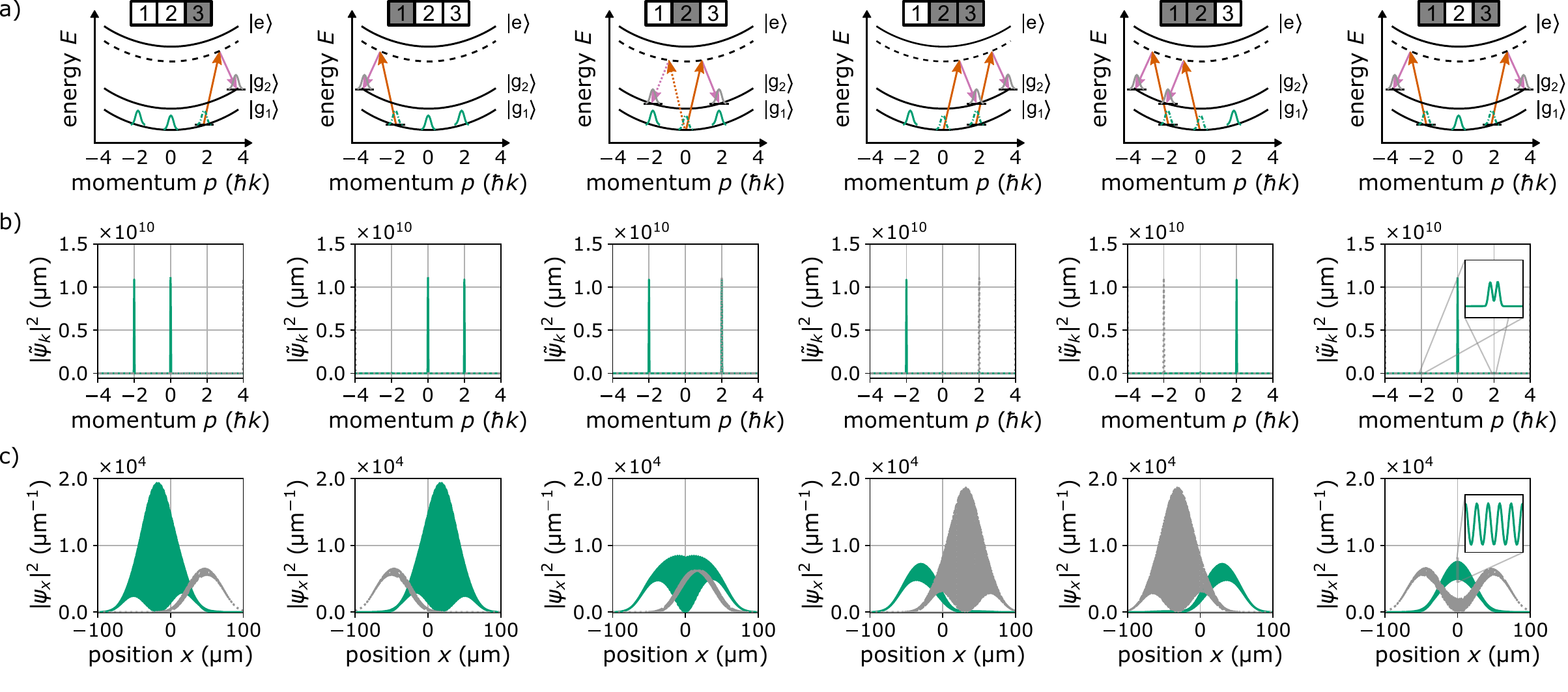}
\caption{Slit configurations with Raman pulses emulating blocking masks. a) Dispersion relation for a 3-level system with ground-state manifold $\ket{\text{g}_1}$, $\ket{\text{g}_2}$ and excited state $\ket{\text{e}}$. Arrows indicate 2-photon Raman transitions utilized for momentum selective transfers of population from the lower (green) to the higher energetic states (grey). These Raman pulses serve as blocking masks, since the detection of the interferometer output will only be sensitive to one internal state. Single Raman pulses are applied to realize the double-slit masks \scalemath{0.95}{\fcolorbox{black}{white}{$1$}\fcolorbox{black}{white}{$2$}\fcolorbox{black}{gray}{$3$}} and \scalemath{0.95}{\fcolorbox{black}{gray}{$1$}\fcolorbox{black}{white}{$2$}\fcolorbox{black}{white}{$3$}} such that one of the interferometer paths is blocked. To prevent a spatial asymmetry of the Sorkin parameter due to mask \scalemath{0.95}{\fcolorbox{black}{white}{$1$}\fcolorbox{black}{gray}{$2$}\fcolorbox{black}{white}{$3$}}, we use a linear combination of two masks with transitions of opposite momentum transfer (solid and dashed arrows). The single-slit masks \scalemath{0.95}{\fcolorbox{black}{white}{$1$}\fcolorbox{black}{gray}{$2$}\fcolorbox{black}{gray}{$3$}}, \scalemath{0.95}{\fcolorbox{black}{gray}{$1$}\fcolorbox{black}{white}{$2$}\fcolorbox{black}{gray}{$3$}} and \scalemath{0.95}{\fcolorbox{black}{gray}{$1$}\fcolorbox{black}{gray}{$2$}\fcolorbox{black}{white}{$3$}} are generated using two sequential Raman pulses such that two interferometer paths are blocked. The trivial three-slit mask \scalemath{0.95}{\fcolorbox{black}{white}{$1$}\fcolorbox{black}{white}{$2$}\fcolorbox{black}{white}{$3$}} is omitted, where no Raman pulse is applied. The corresponding atomic density distributions are shown in b) momentum and c) position space. The zoom insets in \scalemath{0.95}{\fcolorbox{black}{gray}{$1$}\fcolorbox{black}{white}{$2$}\fcolorbox{black}{gray}{$3$}} indicate an imperfect extinction of the blocking mask. Momentum selectivity leads to an incomplete transfer which causes a high-frequency interference pattern in the spatial density distribution of the lower energetic state.} 
\label{fig:masks}
\end{figure*}

\section{Bragg beamsplitter and Raman blocking masks}

We manipulate the external and internal degrees of the atoms using 2-photon transitions mediated by an intermediate state. For example, for the hyperfine structure of the $^{87}$Rb $\text{D}_2$ line, the ground-state manifold $\ket{\text{g}_1}=\ket{5^2\text{S}_{1/2},F=1}$ and $\ket{\text{g}_2}=\ket{5^2\text{S}_{1/2},F=2}$ is coupled via the excited state $\ket{\text{e}}=\ket{5^2\text{P}_{3/2},F=1}$. The single-photon frequencies are detuned from resonance by a global detuning $\Delta$ to prevent population transfer into the intermediate state and thus decoherence due to spontaneous emission. The relative frequency detuning of the two light fields is set relative to the resonance condition
\begin{equation}
    \omega_1-\omega_2=\mathrlap{\overbrace{\phantom{\omega_\text{rec}}}^{\text{Bragg}}}
      \mathrlap{\underbrace{\phantom{\,\omega_\text{rec}+\omega_\text{hfs}+\omega_\text{d}}}_{\text{Raman}}}
    \,\omega_\text{rec}+\omega_\text{hfs}+\omega_\text{d} \text{,} 
\end{equation}
where $\omega_\text{rec}\approx2\hbar k^2 /m$ is the 2-photon recoil frequency, $\omega_\text{hfs}$ is the hyperfine splitting and $\omega_\text{d}$ denotes the Doppler detuning. In the case of Bragg pulses, the internal state is conserved and the detuning is given only by the recoil equal to the kinetic energy gained through the momentum transfer. For the Raman pulses, we include a detuning equal to the energy difference of the two states $\ket{\text{g}_1}$ and $\ket{\text{g}_2}$. In addition, we allow for a Doppler detuning of $\omega_\text{d}=\{0,\pm\omega_\text{rec}\}$ which depends on the momentum $|p\rangle=\{|p_0\rangle, |p_0\pm 2\hbar k\rangle\}$ of the \emph{interferometer paths} $\{\scalemath{0.9}{\fcolorbox{black}{white}{$1\vphantom{3}$}}, \scalemath{0.9}{\fcolorbox{black}{white}{$2$}}, \scalemath{0.9}{\fcolorbox{black}{white}{$3$}}\}$ being addressed. The sub-recoil momentum distribution of the delta-kick collimated \ac{BEC}  ($\sigma_p\ll 2 \hbar k$) enables close-to-unity diffraction fidelities and an exquisite momentum selectivity. These are favorable for our experimental protocol to achieve high extinctions of the blocking masks and diminished cross-talk with off-resonant paths.

Figure~\ref{fig:masks}a) depicts the blocking masks for $3!$ configurations omitting the trivial case of \scalemath{0.9}{\fcolorbox{black}{white}{$1$}\fcolorbox{black}{white}{$2$}\fcolorbox{black}{white}{$3$}} (see figure~\ref{fig:simulation} instead). For the three-slit mask, no Raman pulse is applied such that all wave packets remain in the internal state $\ket{\text{g}_1}$. For the other slit configuration, we illustrate the energy-momentum parabola for the internal states $\ket{\text{g}_1}$, $\ket{\text{g}_2}$, $\ket{\text{e}}$ and indicate the involved 2-photon Raman transitions to emulate the corresponding blocking masks. 

The double-slit configurations \scalemath{0.9}{\fcolorbox{black}{white}{$1$}\fcolorbox{black}{white}{$2$}\fcolorbox{black}{gray}{$3$}}, \scalemath{0.9}{\fcolorbox{black}{gray}{$1$}\fcolorbox{black}{white}{$2$}\fcolorbox{black}{white}{$3$}} and \scalemath{0.9}{\fcolorbox{black}{white}{$1$}\fcolorbox{black}{gray}{$2$}\fcolorbox{black}{white}{$3$}} are realized using a single Raman pulse to transfer population of one of the \emph{interferometer paths} from $\ket{\text{g}_1}$ into $\ket{\text{g}_2}$ under an additional momentum transfer of $\pm2\hbar k$. Controlling the sign of the wave vectors, we choose the direction of transfer towards higher momenta to diminish off-resonant coupling to unwanted paths. In the case of \scalemath{0.9}{\fcolorbox{black}{white}{$1$}\fcolorbox{black}{gray}{$2$}\fcolorbox{black}{white}{$3$}}, the momentum transfer can be directed either way. To prevent a spatial asymmetry of the Sorkin parameter, we use a linear combination of two masks with transitions of opposite momentum transfer. We consider these masks in Eqs.~(\ref{eq:sorkin_epsilon})--(\ref{eq:sorkin_delta}) with weighting factors of 1/2.

The single-slit configurations \scalemath{0.9}{\fcolorbox{black}{white}{$1$}\fcolorbox{black}{gray}{$2$}\fcolorbox{black}{gray}{$3$}}, \scalemath{0.9}{\fcolorbox{black}{gray}{$1$}\fcolorbox{black}{white}{$2$}\fcolorbox{black}{gray}{$3$}} and \scalemath{0.9}{\fcolorbox{black}{gray}{$1$}\fcolorbox{black}{gray}{$2$}\fcolorbox{black}{white}{$3$}} are generated by applying two sequential Raman pulses, such that only a single wave packet remains in the initial state $\ket{\text{g}_1}$. In case of \scalemath{0.9}{\fcolorbox{black}{white}{$1$}\fcolorbox{black}{gray}{$2$}\fcolorbox{black}{gray}{$3$}} and \scalemath{0.9}{\fcolorbox{black}{gray}{$1$}\fcolorbox{black}{gray}{$2$}\fcolorbox{black}{white}{$3$}}, the temporal order of the transitions is chosen such that the Raman pulses address wave packets starting from higher to lower initial momenta.

\begin{figure}[t]
    \begin{center}
    \includegraphics[width=1\columnwidth]{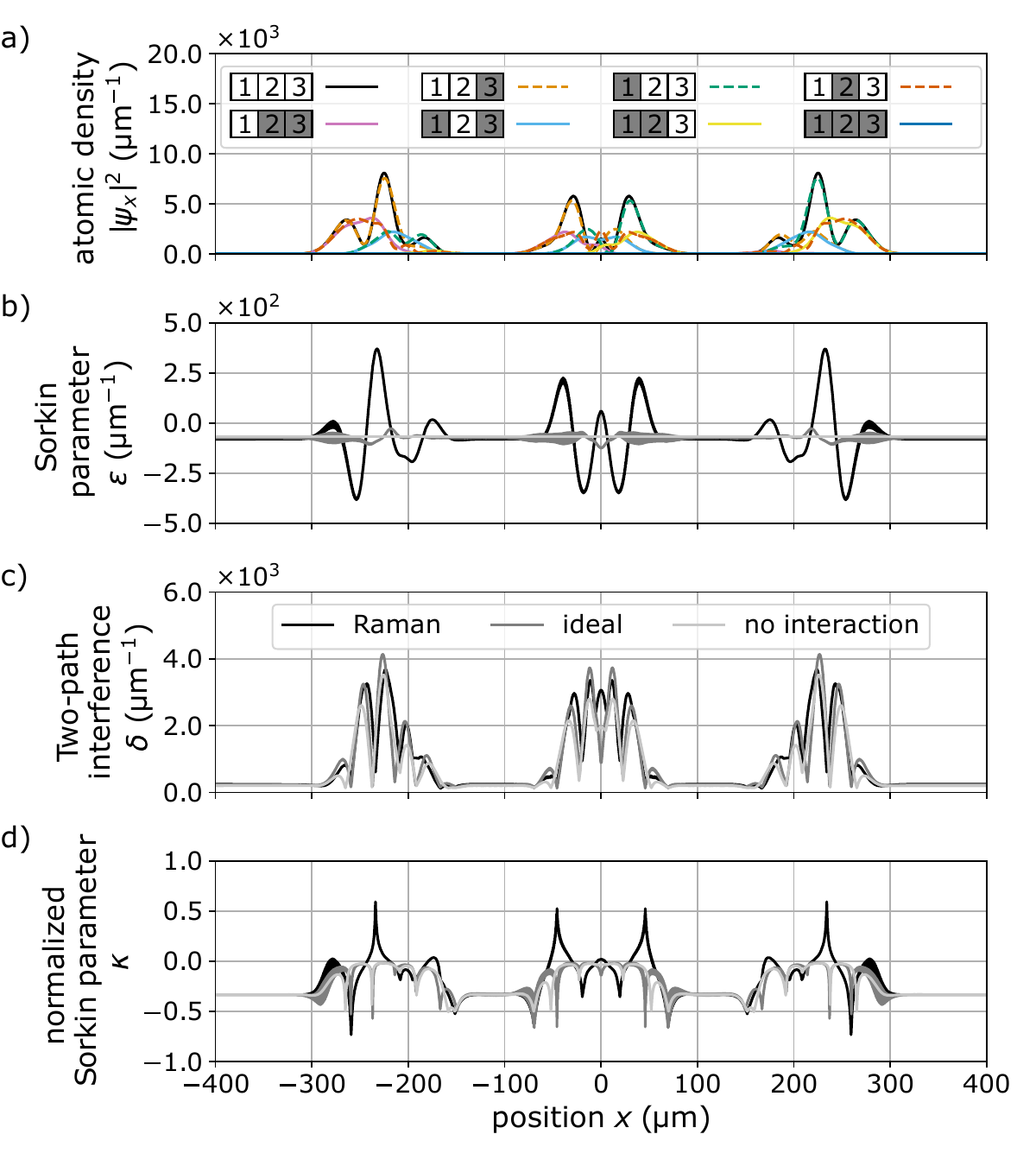}
    \caption{Light-induced multipath interference in a \ac{BEC}. a) Spatial interference pattern for the $3!+1$ slit configurations together with the detector offset. b) Sorkin parameter, c) expected two-path interference and d) normalized Sorkin parameter deduced from Eqs.~(\ref{eq:sorkin_epsilon})--(\ref{eq:sorkin_delta}), respectively. We compare the results obtained with Raman pulses (black) and ideal blocking masks for an interacting (dark grey) and non-interacting (light grey) \ac{BEC}. All three data sets share the same experimental parameters (see text for details). Detector offset, self-interaction and imperfect interferometer pulses lead to a non-vanishing Sorkin parameter.}
    \label{fig:sorkin}
    \end{center}
\end{figure}

\section{Numerical Simulation}

We model the protocol explained above for a self-interacting \ac{BEC} of $^{87}$Rb atoms and simulate the dynamics given by the Gross-Pitaevskii equation \cite{pethick08,pitaevskii03}
using a Python implementation of the Split-step Fourier method \cite{vowe20}. We choose realistic parameters based on the BEC interferometer experiment as reported in Refs.~\cite{abend16,gebbe21}. The simulations are performed on a one-dimensional grid, where we use an effective interaction Hamiltonian $\mathcal{H}_{\text{int}}=g_{\text{1d}}|\psi_{x}|^2$ to match the asymptotic expansion rate of the \ac{BEC} in the simulation with the experiment \cite{kanthak21}. The interaction strength $g_{\text{1d}}=\SI{5e8}{\per\meter\squared}\cdot4\pi\hbar^2 a_{\text{s}} N / m$ is defined by the atom number $N$ and s-wave scattering length, which we set to hundred Bohr radii $a_{\text{s}}=100 a_0$ \cite{harber2002}.  
 
Intermediate steps of our simulation are depicted in figure~\ref{fig:simulation}. In particular, we generate a \ac{BEC} of $N=3\times10^4$ atoms inside a harmonically shaped trapping potential of angular trap frequency $\omega_x = 2\pi\cdot \SI{340}{\hertz}$ and obtain its ground state via the method of imaginary time evolution \cite{goldberg67}. At $t=0$, the \ac{BEC} is released from the trap to propagate freely. Driven by the conversion of interaction into kinetic energy, the condensate expands from its initial size $\sigma_x=\SI{1}{\micro\meter}$ to $\sigma_x=\SI{29}{\micro\meter}$ within $t_\text{exp}=\SI{20}{\milli\second}$ of expansion time. At this point, the initial trap potential is re-applied for a short duration of $\tau_\text{dkc}=\SI{50}{\micro\second}$ causing a reduction in momentum width from $\sigma_p=0.24\hbar k$ to $\sigma_p=0.01\hbar k$ via \ac{DKC}. Here, the size and momentum width of the condensate are given as root-mean-square deviation of the respective density distributions.

\begin{figure*}[t]
    \begin{center}
    \includegraphics[width=2.1\columnwidth]{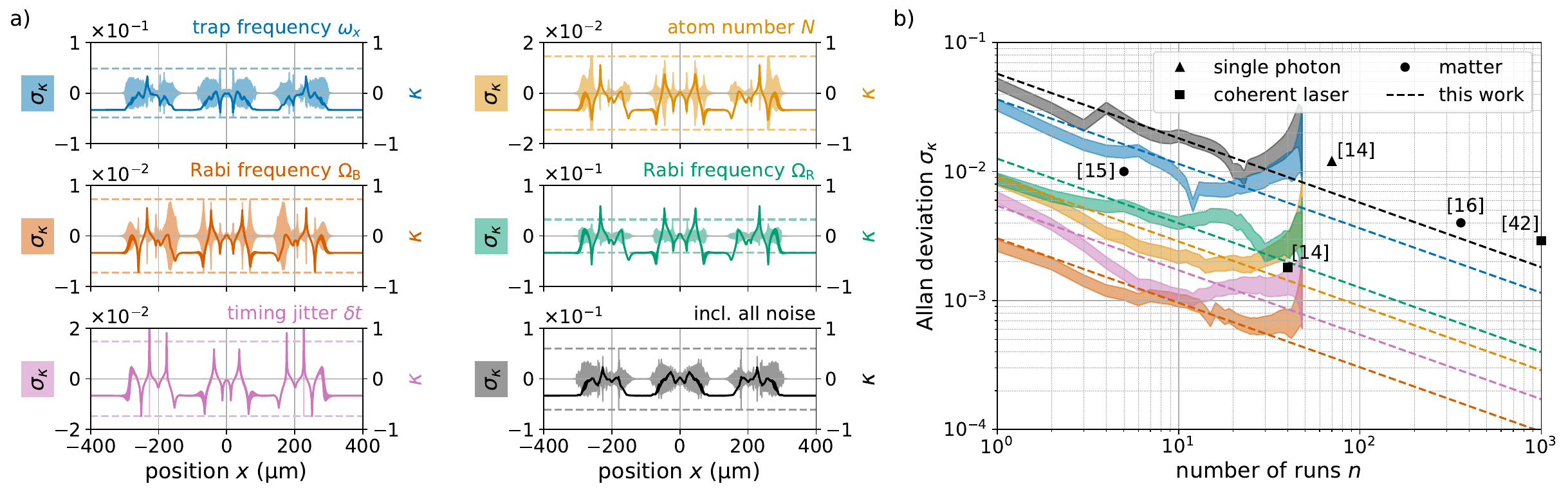}
    \caption{Noise analysis. We study the impact of experimental uncertainties on the outcome of our experimental protocol. a) Selected experimental parameters are subjected to random noise following a normal distribution (see table \ref{tab:ErrorSources} for details). The panels show the mean value of the Sorkin parameter $\kappa$ (solid line, right axis) and its uncertainty $\sigma_\kappa$ (shaded area, left axis) for $n=100$ iterations under variation of individual parameters (colors) and with all parameters varied simultaneously (black). We consider each grid point as a separate experiment and determine an upper bound of $\sigma_\kappa$ as the maximum uncertainty in position space (dashed horizontal lines). b) Overlapping Allan deviation $\sigma_{\kappa}$ averaged over iterations ranging from $1$ to $1000$. The shaded areas indicate the confidence intervals with the same color code as in a). Fits interpolate the datasets of $n=100$ iterations (dashed lines), see text for details. We show our numerical results of the light-pulse atom interferometer in comparison to Born rule tests in literature employing various interferometer sources with macroscopic material slits and blocking masks \cite{sinha10,cotter17,barnea18,soellner12}.} 
    \label{fig:noise}
    \end{center}
\end{figure*}

After $t_\text{del}=\SI{2}{\milli\second}$, we apply the first double Bragg pulse of the interferometer. All light fields are detuned from resonance by $\Delta=\SI{1}{\giga\hertz}$. The 2-photon Rabi frequency is $\Omega_\text{B}=2\pi\cdot\SI{500}{\hertz}$ and we set the pulse duration to $\tau_\text{B}\approx\SI{430}{\micro\second}$ to achieve an equal population in all three desired \emph{interferometer paths} $\{\fcolorbox{black}{white}{$1$}, \fcolorbox{black}{white}{$2$}, \fcolorbox{black}{white}{$3$}\}$. The light fields are modeled as Gaussian beams of waist $w_0=\SI{3.1}{\milli\meter}$ such that the Rayleigh range $kw_0^2/2\gg \sigma_x$. The wave packets then separate for $t_\text{int}=\SI{2}{\milli\second}$, before we redistribute them among the three \emph{interferomter output ports} via the second identical double Bragg pulse. 

Symmetrically centered between the two Bragg pulses, we apply the blocking masks via the respective single Raman pulses with $\Omega_\text{R}=2\pi\cdot\SI{2}{\kilo\hertz}$ and $\tau_\text{R}=\SI{250}{\micro\second}$ adjusted for a complete population transfer. Doppler-detuning and sign of wave vectors are chosen according to the selected \emph{interferometer paths} and direction of momentum transfers. To account for the additional time passing during the Raman pulses, we include a free evolution of $2\tau_{\text{R}}$ for the three-slit mask (no pulse) and $\tau_{\text{R}}$  for the double-slit masks (one pulse). This is done to keep the total evolution time the same as for the single-slit masks (two pulses). Figure~\ref{fig:masks} shows the atomic density distribution of the ground-state manifold in b) momentum and c) position space. Shortly after the Raman pulses, the wave packets of the \emph{interferometer paths} spatially overlap, which leads to high-frequency interference of different momentum components within the same internal state. The zoom insets in the \scalemath{0.9}{\fcolorbox{black}{gray}{$1$}\fcolorbox{black}{white}{$2$}\fcolorbox{black}{gray}{$3$}} indicate an imperfect extinction of the respective blocking masks. An incomplete population transfer, i.e. due to momentum selectivity of the Raman pulses imposed by the limited Fourier width of the light pulses in combination with a non-vanishing expansion of the atomic ensemble, leads to the interference visible in the lower energetic state.

After a separation of $t_\text{sep}=\SI{18}{\milli\second}$, we directly access the interference pattern in the spatial density distribution from the \emph{interferometer output ports} of each slit configuration. From these, the Sorkin parameter $\epsilon$, expected two-path interference $\delta$ and normalized Sorkin parameter $\kappa$ are determined according to Eqs.~(\ref{eq:sorkin_epsilon})--(\ref{eq:sorkin_delta}). We include an offset $\rho_{_{\,\mathrm{\scalemath{0.9}{  \fcolorbox{black}{gray}{$_1$}\fcolorbox{black}{gray}{$_2$}\fcolorbox{black}{gray}{$_3$} }}}}=\text{max}\left(|\psi_x|^2\right)/\text{SNR}$ to account for inevitable detector noise and background signals assuming a signal-to-noise ratio of $\text{SNR}=100$.

Figure~\ref{fig:sorkin}a) shows the interference pattern for various slit configurations together with the detector offset (colored lines). The Sorkin parameter of Eq.~(\ref{eq:sorkin_epsilon}), the two-path interference and the normalized Sorkin parameter of Eq.~(\ref{eq:sorkin_kappa}) are shown in figures~\ref{fig:sorkin}b)--d). We compare the outcome for the scenarios of no interaction (light grey), ideal blocking masks (dark grey) and Raman pulses as blocking masks (black). In the case of the ideal masks, we block a momentum state $|p\rangle$ in momentum space by simply setting the wave function between $p\pm\hbar k$ to zero. The non-interacting case employs ideal masks with $g_{\text{1d}}=0$.

As expected for the absence of interaction, $\epsilon$ basically vanishes and is only shifted from zero by the detector offset $\rho_{_{\,\mathrm{\scalemath{0.9}{  \fcolorbox{black}{gray}{$_1$}\fcolorbox{black}{gray}{$_2$}\fcolorbox{black}{gray}{$_3$} }}}}$. Self-interaction then leads to a non-vanishing $\epsilon$ including residual interference from non-linear phase shifts visible in $\delta$. Further deviations from zero then occur due to imperfect Raman pulses caused by incomplete population transfers and coupling to off-resonant paths. For all scenarios, $\epsilon$ and $\delta$ approach $-\rho_{_{\,\mathrm{\scalemath{0.9}{  \fcolorbox{black}{gray}{$_1$}\fcolorbox{black}{gray}{$_2$}\fcolorbox{black}{gray}{$_3$} }}}}$ and $3\rho_{_{\,\mathrm{\scalemath{0.9}{  \fcolorbox{black}{gray}{$_1$}\fcolorbox{black}{gray}{$_2$}\fcolorbox{black}{gray}{$_3$} }}}}$ for vanishing atomic densities. This is reflected in $\kappa$ as an overall offset of $-1/3$, in accordance with Eqs.~(\ref{eq:sorkin_epsilon})--(\ref{eq:sorkin_delta}). 

The obtained values of $\epsilon$ and $\kappa$ are estimates of the expected outcome of the Born rule test for the given experimental protocol and parameter set and will serve as reference for the following noise examinations.

\section{Noise analysis}

We examine the impact of experimental uncertainties on the outcome of our experimental protocol. Selected parameters cover the atom source, light pulses and interferometer itself, which we subject to random noise following a normal distribution. We assume slowly varying parameters meaning that a parameter is kept constant for a complete set of masks before being changed for the next measurement run. For the atom source, we focus on the variation of the \ac{BEC}'s momentum distribution, which is imposed by the angular trap frequency $\omega_x$ of the \ac{DKC} potential, and fluctuations in the number of atoms $N$. The light pulses of the interferometer are subjected to a variation in Rabi frequencies $\Omega_\text{R}$ and $\Omega_\text{B}$, i.e. imposed by intensity drifts of the generating light fields. Additionally, we include a timing jitter $\delta t$ to the interferometer time $t_\text{int}$. Table \ref{tab:ErrorSources} shows the mean values $\xi$ and standard deviations $\sigma_{\xi}$ of the parameters under examination.

\begin{table}[b]
    \caption{Contribution of the experimental uncertainties to the statistical deviation $\sigma_\kappa$ of the Sorkin parameter $\kappa$.}
    \centering
    \begin{tabular}{l|c|c}
        \hline
        \hline
        parameter $\xi$ & $\xi\pm\sigma_{\xi}$ & $\sigma_{\kappa}\left(100\,\text{runs}\right)$ \\
        \hline
        \textcolor{seaborn1}{trap frequency} $\omega_x$ & $2\pi\cdot(340\pm10)\,\si{\hertz}^{\ddagger}$ & \transparent{0.3}\fcolorbox{seaborn1}{seaborn1}{\transparent{1}\SI{5e-2}{}} \\ 
        \textcolor{seaborn2}{atom number} $N$ & $(3.0\pm0.5)\times 10^4$ & \transparent{0.3}\fcolorbox{seaborn2}{seaborn2}{\transparent{1}\SI{1e-2}{}} \\ 
        \textcolor{seaborn3}{Rabi frequency} $\Omega_{\text{R}}$ & $2\pi\cdot(2.0\pm0.2)\,\si{\kilo\hertz}$ & \transparent{0.3}\fcolorbox{seaborn3}{seaborn3}{\transparent{1}\SI{3e-2}{}}\\
        \textcolor{seaborn4}{Rabi frequency} $\Omega_{\text{B}}$ & $2\pi\cdot(500\pm50)\,\si{\hertz}$ & \transparent{0.3}\fcolorbox{seaborn4}{seaborn4}{\transparent{1}\SI{7e-3}{}}\\
        \textcolor{seaborn5}{timing jitter} $\delta t$ & $(10^4\pm5)\,\si{\nano\second}$ & \transparent{0.3}\fcolorbox{seaborn5}{seaborn5}{\transparent{1}\SI{1e-2}{}} \\
        \hline 
        all of the above & & \transparent{0.3}\fcolorbox{black}{black}{\transparent{1}\SI{6e-2}{}} \\
        \hline
        \hline
        \multicolumn{3}{l}{$^{\ddagger}$\footnotesize{The corresponding momentum width is $\sigma_p = (0.013${\raisebox{0.5ex}{\tiny$^{-0.008}_{+0.013}$}}$)\hbar k$.}} \\
    \end{tabular}
    \label{tab:ErrorSources}
\end{table}

For comparison of the individual noise contributions, we simulate the experimental protocol as stated in the previous section for $n=100$ iterations with a single parameter being varied, while the others are fixed to their respective mean value. In the case of the timing jitter, we add an extra $\delta t = \SI{10}{\micro\second}$ to the interferometer time. For the benchmark of our experimental protocol, we run the simulation $n=100$ times with all parameters being changed simultaneously using the same seed for the randomly generated values. 

Figure~\ref{fig:noise}a) shows the mean value of the Sorkin parameter $\kappa$ (solid line, right axis) and its uncertainty $\sigma_\kappa$ (shaded area, left axis) for $n=100$ iterations under variation of individual parameters (colors) and with all parameters varied simultaneously (black). The uncertainty is taken as the the standard deviation divided by the square root of the number of measurements \cite{barnea18}. We consider each grid point as a separate experiment and determine an upper bound of $\sigma_\kappa$ as the maximum uncertainty in position space (dashed horizontal lines). The statistical deviations determined in this way are given in table \ref{tab:ErrorSources} as $\sigma_{\kappa}\left(100\,\text{runs}\right)$.

Figure~\ref{fig:noise}b) displays the overlapping Allan deviation $\sigma_{\kappa}$ averaged over the number of iterations ranging from 1 to 1000. The shaded areas indicate the confidence intervals under simultaneous variation of all parameters (black) and with single varying parameters (colors, with same color code as in figure~\ref{fig:noise}a) and table \ref{tab:ErrorSources}).  Fits of the form  $\sigma_{\kappa}=\text{constant}\,/\sqrt{n}$ reflect the expected scaling of the standard deviation with the square root of runs (dashed lines), interpolating the datsets of $n=100$ iterations. In addition, we show our numerical results in comparison with previously published Born rule tests (data points) employing various interferometer sources with macroscopic material slits and blocking masks \cite{sinha10,cotter17,barnea18,soellner12}.

\section{Discussion}

We have proposed light-pulse atom interferometry with \acp{BEC} as an indirect test of Born’s rule. Assuming that the modulo-square rule holds, we compute a non-zero Sorkin parameter with an upper bound of $\sigma_{\kappa}(100\,\text{runs}) = 5.7\times10^{-3}$ and $\sigma_{\kappa}(1000\,\text{runs}) = 1.8\times10^{-3}$ on its statistical deviation for a realistic experimental configuration. In an experimental implementation of our specific protocol with its key parameters controlled as stated in table \ref{tab:ErrorSources}, any deviation beyond this bound could hint towards a violation of Born's rule.

For the selected parameter set, the deviation from the non-zero Sorkin parameter indicated in figure~\ref{fig:noise}a) mainly results from unsteady population transfers of the Raman pulses due to the fluctuations in momentum width $\sigma_p = (0.013${\raisebox{0.5ex}{\tiny$^{-0.008}_{+0.013}$}}$)\hbar k$ and Rabi frequency. For example, for the \scalemath{0.9}{\fcolorbox{black}{white}{$1$}\fcolorbox{black}{white}{$2$}\fcolorbox{black}{gray}{$3$}} mask, the fidelities of the Raman pulses vary by $5\%$ and $3\%$, respectively. Non-linear contributions stemming from self-interaction of the \ac{BEC} are imposed by the atomic density and its distribution among the interferometer paths via the Bragg pulses. These are prominently suppressed by one order magnitude given by the extended expansion times (linear regime). The assumed timing jitter introduced in the interferometer is typical for state-of-the-art pulse generators, but has only a minor impact on the deviation for the selected interferometer times.

As displayed in the benchmark in figure~\ref{fig:noise}b), our method is competitive with previous Born rule tests. Collecting the necessary experimental data seems feasible with recent progress in rapid \ac{BEC} production \cite{rudolph15,herbst22}, bringing a Born rule test with experimental cycle times of smaller than \SI{10}{\second} in reach. Using optical diffraction lattices instead of material slits and blocking masks avoids possible systematic errors due to geometrical inaccuracies from the manufacturing process and allows to flexibly control the parameter set of the multipath interferometer. In addition, imperfection of our Raman blocking masks are accessible via detection of the higher energetic state $\ket{\text{g}_2}$, which could be utilized in a post-correction of the Sorkin parameter.

Our simulations are not restricted to one dimension and could in principle be extended to a more realistic experiment including light-field distortions \cite{neumann21,fitzek20} as well as temporal envelopes and frequency chirps of the optical pulses. In fact, it has been shown that adiabatic-rapid passages in combination with diffraction pulses are more robust against fluctuations of the pulse intensity and detuning \cite{kovachy12} and could ease the requirements on the stability of the experimental parameters.  We further anticipate optimization of the parameter sets by means of a machine-learning algorithm to make the Bragg beamsplitters and Raman pulses less susceptible to experimental noise \cite{li24}.

\section{Acknowledgements}

\noindent We would like to thank the QUANTUS collaboration for fruitful discussions. This work is supported by the German Space Agency (DLR) with funds provided by the Federal Ministry for Economic Affairs and Climate Action (BMWK) under grant number 50WM1953 (SK, JP and MK), 50WM2250B (SK and MK) and 50MW2251 (DR and MK).\\
\newline

\bibliography{bibliography}
\bibliographystyle{prstytitlenew}

\appendix
\section{Appendix}

The simulations are performed on one-dimensional position and momentum grids of size $2^{16}$ ranging between $\left[-x_{\text{max}},+x_{\text{max}}\right]$ and $\left[-p_{\text{max}},+p_{\text{max}}\right]$, respectively. We set the size of the position grid to $x_{\text{max}}=\SI{1}{\milli\meter}$, which is sufficiently large to accommodate the wave packets at the end of our experimental protocol. In particular, the separation of the wave packets in the $j^\text{th}$ diffraction order is $ j\cdot  2\hbar k/m \cdot (t_\text{int}+t_\text{sep})=j\cdot \SI{235}{\micro\meter}$ with a final size of $\sigma_x=\SI{31}{\micro\meter}$. The spatial resolution $\mathrm{d}x=\SI{31}{\nano\meter}=\SI{4e-2}{}\lambda$ is capable to resolve the atomic dynamics in the optical lattice of wavelength $\lambda=\SI{780}{\nano\meter}$ \cite{fitzek20}. The momental resolution $\mathrm{d}p=\SI{4e-4}{}\hbar k$ is sufficiently small to resolve the momentum width $\sigma_p=\SI{1e-2}{}\hbar k$ of the delta-kick collimated \ac{BEC}. The computed size of the momentum grid is $p_{\text{max}}=12.6\hbar k$, which in principle allows to represent momentum states up to the $6^\text{th}$ diffraction order. However, the grid's spatial extent sets the upper limit to the $4^\text{th}$ diffraction order with reasonable truncation of the wave packets during detection. This provides $\pm4\hbar k$ of momenta for our interferometric sequence and an additional $\pm4\hbar k$ for potential residual populations of higher orders during the atom-light interaction.

The typical time scales of the dynamics differ during the experimental sequence and are set by the respective kinetic energies of the wave packets and the external potentials applied. To account for this, convergence studies are carried out to determine the time resolution needed for the intermediate steps of our protocol. For example, the time steps during the interferometer pulses are $\mathrm{d}t=\tau/100$, which is on the order of $\SI{1}{\micro\second}$ but can be as large as $\SI{10}{\milli\second}$ for the free evolution prior detection.

\end{document}